\documentclass[11pt]{amsart}
\usepackage{hyperref}

\theoremstyle{definition}

\theoremstyle{remark}

\numberwithin{equation}{section}

\def \dd{{\rm d}}
\def \DD{{\rm D}}

\begin{document}
\title[]{Reconsidering Schwarzschild's original solution}%
\author{Salvatore Antoci}
\address{Dipartimento di Fisica ``A. Volta'' and INFM, Pavia, Italy}
\email{Antoci@fisicavolta.unipv.it}%
\author{Dierck Ekkehard Liebscher}
\address{Astrophysikalisches Institut Potsdam, Potsdam, Germany}
\email{deliebscher@aip.de}%
\thanks{}%
\begin{abstract}
We analyse the Schwarzschild solution in the context of the
historical development of its present use,
and explain the invariant definition of the Schwarzschild's radius
as a singular surface, that can be applied
to the Kerr-Newman solution too.
\end{abstract}
\maketitle
\section{Introduction: Schwarzschild's solution and the ``Schwarzschild'' solution}
Nowadays simply talking about Schwarzschild's solution requires a
preliminary reassessment of the historical record as {\em conditio
sine qua non} for avoiding any misunderstanding. In fact, the
present-day reader must be firstly made aware of this seemingly
peculiar circumstance: Schwarzschild's spherically symmetric,
static solution \cite{Schw16a} to the field equations of the
version of the theory proposed by Einstein \cite{Einstein15a} at
the beginning of November 1915 is different from the
``Schwarzschild'' solution that is quoted in all the textbooks and
in all the research papers. The latter, that will be here always
mentioned with quotation marks, was found by Droste, Hilbert and
Weyl, who worked instead \cite{Droste17}, \cite{Hilbert17},
\cite{Weyl17} by starting from the last version \cite{Einstein15b}
of Einstein's theory.\footnote{Recent revisions of the relevant documents
tell that Schwarzschild at that time was only contacting Einstein
about the question of the solution \cite{Treder01}.}
As far as the vacuum is concerned, the two
versions have identical field equations; they differ only because
of the supplementary condition
\begin{equation}\label{1}
\det{(g_{ik})}=-1
\end{equation}
that, in the theory of November 11th, limited the covariance to the
group of unimodular coordinate transformations. Due to this
fortuitous circumstance, Schwarzschild could not simplify his
calculations by the choice of the radial coordinate made e.g. by
Hilbert; he was instead forced to adopt ``polar coordinates with determinant
1'' that led him to a solution depending on two parameters instead
of the single one found by Hilbert. Schwarzschild could
then fix one of the parameters in such a way as to push in the
``Nullpunkt'' the singularity that is named after him. This move is
impossible with Hilbert's choice, and this is the origin of the
difference between Schwarzschild's solution and the
``Schwarzschild'' solution of the literature which, to put an end
to the present confusion, should be more aptly named after Hilbert.

\section{Solving Schwarzschild's problem without fixing the radial
coordinate}
    In order to display the difference between the
spherically symmetric, static solutions found with different
choices for the radial coordinate, let us briefly recall the
calculation \cite{Combridge23}, \cite{Janne23} done by Combridge
and by Janne without fixing the radial coordinate at all. By
following de Sitter \cite{de Sitter16}, they found that the line
element of the most general static, spherically symmetric field
can be reduced by choice of adapted coordinates to the form
\begin{equation}\label{2}
\dd s^2=-\exp{\lambda}\dd r^2
-\exp{\mu}[r^2(\dd\vartheta^2+\sin^2{\vartheta}\dd\varphi^2)]
+\exp{\nu}\dd t^2
\end{equation}
in symmetry adapted coordinates. Here $\lambda$, $\mu$, $\nu$ are
functions of $r$ only. The proof of the correctness of de Sitter's
choice can be found e.g. in Eiesland's paper \cite{Eiesland25}. By
availing of the field equations already made explicit \cite{de
Sitter16} by de Sitter, the mentioned authors found that a
solution that is Minkowskian at the spatial infinity could be
expressed in terms of one arbitrary function $f(r)$ and of its
derivative ${f'}(r)$ by setting:
\begin{eqnarray}\label{3}
\exp{\lambda}=\frac{{f'}^2}{1-2m/f},\\\label{4}
\exp{\mu}=\frac{f^2}{r^2},\\\label{5}
\exp{\nu}=1-2m/f,
\end{eqnarray}
where $m$ is a constant, while of course the arbitrary function
$f$ must have the appropriate behaviour as $r\rightarrow\infty$.
It is now an easy matter to reproduce the results that one obtains
when the most popular coordinate conditions are imposed. With the
exception of Schwarzschild's case, the latter were adopted with the
aim of making the calculations easier. However, a simple glance to
the papers by Combridge and Janne shows that the calculations were
not as {\em extra vires} as to be in urgent need of
simplification.\par Setting $\lambda=0$ provides Droste's initial
way of fixing the radial coordinate, as given in his equation (4).
Despite this initial choice, the ``Schwarzschild'' solution made
its first appearance just in Droste's paper \footnote{Insisting
with the coordinate condition $\lambda=0$ leads one to define $f$
through the hardly solvable equation
\begin{eqnarray}\nonumber
r=f(1-2m/f)^{1/2}
+m\ln{\frac{f^{1/2}+(f-2m)^{1/2}}{f^{1/2}-(f-2m)^{1/2}}}
+ {\rm const}.
\end{eqnarray}
This is why Droste changed horses {\em in itinere} and eventually
succeeded in becoming the forerunner of Hilbert.}.\par Choosing
$\mu=0$ leads instead straight to the ``Schwarzschild'' solution,
i.e. to Hilbert's form \cite{Hilbert17}, given by equation (45) of
his monumental paper. One notes that if one attributes to $r$ the
usual range $0\leq r<\infty$ equation (\ref{5}) would only admit
the value $m=0$ of the mass constant. Maybe this is the reason why
Hilbert deviated from de Sitter's choice (\ref{2}) for the line
element and dropped the {\em a priori} condition of persistence of
the sign for the components of the metric. The omission allows for
a nontrivial solution, but also for the appearance of the well
known singularity of the components of the metric at the
``Schwarzschild'' radius. The singularity separates an outer
region $r>2m$, that with the lapse of the years proved itself
capable to precisely account for the workings of Nature, from an
inner region whose occurrence soon evolved from a slightly
annoying peculiarity into a major conundrum
\cite{Brillouin23}.\par Posing $\lambda=\mu$ produces the solution
expressed in isotropic coordinates, ingeniously found by Weyl
through a coordinate transformation \cite{Weyl17} by starting from
his own ``Schwarzschild'' solution. Here again we get a singular
behaviour, this time due to the vanishing of $\det(g_{ik})$ when
$r=m/2$. There is no track here of the tantalizing inner region of
Hilbert. With isotropic coordinates Nature is well accounted for,
maybe too well. There is in fact a little redundance: only the
outer part of Hilbert's solution is allowed to appear, but it
appears twice, once  for $m/2<r<\infty$, and a second time in the
range $0<r<m/2$. The two replicas happen to be joined just at the
``Schwarzschild'' singularity. Confronted with such {\em embarras
de richesse} Weyl flatly declared that in Nature only ``ein
St\"uck'' of one of the two copies, not reaching the singularity,
must be realised \cite{Weyl17}. Einstein and Rosen \cite{ER35}
used the full structure to derive an argument for the mass to be
positive. They felt one could postulate the necessity of hiding
the curvature singularity by constructing such a ``bridge'' and
identify it with a particle. Their hope to get a hint at some kind
of quantisation was not met, but they found that only a positive
mass allows to find that bridge.\par Schwarzschild's true and
authentic solution \cite{Schw16a}, though written with the usual
polar coordinates rather than with the original ``polar
coordinates of determinant 1'', can instead be retrieved by
imposing the condition $\lambda+2\mu+\nu=0$. Due to (\ref{3}),
(\ref{4}) and (\ref{5}) $f$ must then fulfil the equation
\begin{equation}\label{6}
\frac{{f'}^2f^4}{r^4}=1,
\end{equation}
i.e., with appropriate choice of sign, ${f'}=r^2/f^2$. The latter
equation integrates to
\begin{equation}\label{7}
f=(r^3+\varrho)^{1/3}
\end{equation}
where $\varrho$ is a second integration constant, besides $m$. Had
not he died just a few months after his discovery due to a rare
illness contracted while at war on the Russian front, one could
say that Schwarzschild, besides being very clever, was also a very
lucky man. In fact the burden of the coordinate condition
(\ref{1}) he was {\em obtorto collo} forced to confront by the
version of Einstein's theory he was aware of turned out to be a
blessing, when compared to the simplifying assumptions adopted by
the later authors, that could enjoy the eventually conquered
general covariance \cite{Einstein15b}, \cite{Hilbert15}. Although
the move was later declared not recommendable by Hilbert with a
footnote of devastating authority that decided the destiny of the
true Schwarzschild's solution \footnote{See footnote 1 at page 71
of \cite{Hilbert17}.}, with his extra parameter $\varrho$
Schwarzschild had the chance of imposing the continuity of the
components of $g_{ik}$ in the range $0<r<\infty$. He did so by
setting
\begin{equation}\label{8}
\varrho=(2m)^3
\end{equation}
thereby dispatching in the ``Nullpunkt'' what the posterity
would have unanimously called ``the singularity at the
Schwarzschild radius''.

\section{The role of hyperbolic motion in choosing Schwarzschild's solution}
Schwarzschild's position (\ref{8}) is sufficient for complying
with de Sitter's prescription (\ref{2}) for the line element, but
it is by no means necessary. Larger values of $\varrho$ fulfil the
prescription as well; moreover, the solutions with $\varrho\geq (2m)^3$
are different from each other, as the simple consideration of the
maximum value attained by the scalar curvature in each of them
immediately shows. One needs an additional postulate for fixing
the value of $\varrho$. Abrams has shown \cite{Abrams89}
that one can avail of an assumption that seems
to be quite appropriate both from a geometrical
and from a physical standpoint. Let us consider a test body whose
four-velocity is $u^i$; its acceleration four-vector is defined as
\begin{equation}\label{9}
a^i\equiv\frac{\DD u^i}{\dd s}\equiv\frac{\dd u^i}{\dd s}+\Gamma^i_{kl}u^ku^l,
\end{equation}
where $\DD/\dd s$ indicates the absolute derivative. From it one builds
the scalar quantity
\begin{equation}\label{10}
\alpha=(-a_ia^i)^{1/2}.
\end{equation}
The motion of a test body kept at rest in the static, spherically
symmetric field whose line element, in adapted coordinates, is
given by equations (\ref{3}), (\ref{4}) and (\ref{5}), is defined
by postulating the constancy of $r$, $\vartheta$ and $\varphi$. At
first glance, the definition of the world lines of {\em rest}
seems to depend on the particular coordinates we use. However, we can
identify the congruence of world-lines of our {\em particles at
rest} through use of the Killing vectors of the metric. There is
only one time-like Killing congruence that has not only the
Killing property, but is hypersurface orthogonal
($\xi_{[k,l}\xi_{m]} = 0$) too. It is the congruence we identify
in our coordinates with $r,\vartheta,\varphi$ constant. It obeys the
differential equations
\begin{equation}\label{11}
\frac{\DD}{\dd s}\left(\frac{a^i}{\alpha}\right)-\alpha u^i=0,~~~
\alpha={\rm const};
\end{equation}
therefore the test body under question
describes an invariantly defined motion \cite{Rindler60}, that
Rindler aptly called hyperbolic.
The only nonvanishing component
of $a^i$ in a field with the line element (\ref{2}) is
\begin{equation}\label{12}
a^1=\frac{1}{2}\nu'\exp{(-\lambda)}.
\end{equation}
For the solutions given by (\ref{3}), (\ref{4}) and (\ref{5})
the constant $\alpha$ has the expression
\begin{equation}\label{13}
\alpha=\left[\frac{m^2}{f^3(f-2m)}\right]^{1/2}
\end{equation}
in terms of the mass constant $m$ and of the arbitrary function $f$.
As noticed by Rindler, besides the geometrical meaning, $\alpha$
has an immediate physical meaning. Let us consider
a locally Minkowskian coordinate system whose spatial coordinates
be centered at the position $r=r_0$, $\vartheta=\vartheta_0$,
$\varphi=\varphi_0$;
the quantity $\alpha$ equals the strength of the gravitational
pull measured by a dynamometer that holds a unit mass at
rest at the given position.\par
By substituting (\ref{7}) in (\ref{13}) one sees that,
with a fixed value of the mass constant, the maximum
value of the force that can be measured by the dynamometer
is different for different values of $\varrho$. As already
noticed, the solutions are geometrically and physically
different, and we need some reason for choosing one of them.
Abrams' argument is the following \cite{Abrams89}:
since up to now no experimental evidence has been found
for attributing a finite limiting value to $\alpha$, we
cannot help mimicking in the case of general relativity
the way out adopted in Newtonian physics.
There the norm of the force exerted by the gravitational field
of an ideal pointlike mass on a test body tends to
infinity as the test body is brought nearer and nearer to the
source of the field. According to equation (\ref{13}), in the
static spherically symmetric field defined by (\ref{3}),
(\ref{4}) and (\ref{5}) $\alpha\rightarrow\infty$ only
when $f\rightarrow 2m$, because $f\rightarrow 0$ is
prohibited by equation (\ref{5}). If one chooses the
arbitrary function $f$ according to (\ref{7}),
$\alpha$ is allowed to grow without limit at $r=0$ only when
$\varrho$ is chosen just in the way kept by Schwarzschild in
his fundamental paper of 1916
\footnote{To dispel any possible misunderstanding, it is stressed
that no meaning is attributed to the accidental vanishing of
the radial coordinate where $\alpha\rightarrow\infty$.}.

We add a remark on the Killing congruences that shows how to
generalize the argument for the full Kerr-Newman-solution.
The elements
\begin{equation}
\xi^k \frac{\partial}{\partial x^k} = \lambda \frac{\partial}{\partial t}
+ \mu \frac{\partial}{\partial \varphi}
\end{equation}
of the Killing group ($\xi_{k;l}+\xi_{l;k}=0$)
of the Kerr-Newman metric
\begin{eqnarray}
\dd s^2 &=& \frac{\varrho^2}{\Delta}\dd r^2 + \varrho^2\dd\vartheta^2
+\frac{\sin^2\vartheta}{\varrho^2} ((r^2+J^2)\dd\varphi-J\dd t)^2\nonumber\\
&-&\frac{\Delta}{\varrho^2}(\dd t - J\sin^2\vartheta\dd\varphi)^2\\
\Delta &=& r^2+J^2+Q^2-2Mr\nonumber\\
\varrho^2 &=& r^2+J^2\cos^2\vartheta\nonumber
\end{eqnarray}
define invariantly a set of orbits. The acceleration on these orbits
\begin{eqnarray}
\alpha^2 &=& - g_{ij}a^ia^j = - g_{ij}
(\frac{\xi^i}{N})_{;k}\frac{\xi^k}{N}
(\frac{\xi^j}{N})_{;l}\frac{\xi^l}{N}\ ,\\
N&=&\sqrt{-g_{mn}\xi^m\xi^n}\nonumber
\end{eqnarray}
contains always the factor $1/\sqrt{\Delta}$ and diverges for orbits
on the surface $\Delta = 0$. All Killing congruences are spacelike
at $\Delta = 0$ except for the case given by $$\mu = \lambda J/(r_0^2+J^2)\ .$$
This congruence is timelike for $r > r_0 = M + \sqrt{M^2-J^2-Q^2}$
and null on the surface
$r = r_0 = M + \sqrt{M^2-J^2-Q^2}$.\footnote{The physical interpretation
of the congruence is a swarm of particles kept at rest
in the field. This rest, in the static case ($J=0$), is just no change
in the spatial coordinates. In the general case, this ``rest frame''
undergoes a drag by the rotation of the source. The drag is given by $\mu \neq 0$.}
Its acceleration diverges in the limit $r \rightarrow r_0$.
In the case of a static metric, $J=0$, the congruence turns out to be
the hypersurface-orthogonal one.
Hence, the surface $r = r_0 = M + \sqrt{M^2-J^2-Q^2}$
is singular in the Kerr-Newman case, too.
Any Killing congruence that remains timelike in the outer vicinity will
show the singularity.

One can see that the singularity in the acceleration is connected
with the norm of the Killing vector becoming zero or the
determinant of the metric in stationary coordinates becoming zero.
Let us use the notation $\Xi = \xi_a\xi^a$. The velocity along the
orbits is given by $u^k = \xi^k/\sqrt{\Xi}$, the acceleration
$a^k$ by $\Xi^2 a^k = \Xi\ {\xi^k}_{;l}\xi^l -
\xi^k\xi^l\xi_b{\xi^b}_{;l}$. The norm of this expression is $$
\Xi^4\ a_ka^k = \frac{1}{4}\Xi^2\Xi_{,k}\Xi_{,l}g^{kl} -
\Xi(\Xi_{,l}\xi^l)^2 $$ We may, of course, introduce coordinates
in which $\xi^k = \delta^k_0$. These coordinates show that
$\Xi_{,l}\xi^l=g_{00,0}=0$. We obtain $$ a_ka^k =
\frac{1}{4\Xi^2}\Xi_{,k}\Xi_{,l}g^{kl} $$ When $\Xi$ has a zero of
some order n on some surface of orbits, and the space-part metric
has no singularity, the norm of the acceleration has to show a
second-order infinity. If the space-part metric has a singularity,
but the determinant of the metric remains finite, as in the
Schwarzschild case, the infinity of $a_ka^k$ is still of first
order. If $a_ka^k$ becomes infinite while $\Xi$ has no zero, it
must be due to a zero of the determinant of the metric tensor in
the coordinates defined.\footnote{When singular coordinate
transformations are forbidden, the zeros of the determinant of the
metric tensor acquire a physical interpretation \cite{Kreisel67}.
In addition, the Schwarzschild mass can be interpreted as result
of the gravitational field outside the Schwarzschild sphere
\cite{Treder75,Treder78}.}

\section{The analytic extensions of the ``Schwarzschild'' solution}
As previously shown, Hilbert's solution was born out of the
accidental choice of the radial coordinate $r$ produced by setting
$\mu=0$ in equation (\ref{4}); hence there is no reason to accept
as unavoidable consequences of the very field equations of general
relativity all the features stemming from this choice, in
particular the existence of the region for $r<2m$. However
Hilbert's solution was soon perceived as the unique
``Schwarzschild'' solution and as such it became the obligatory
starting point of all the theoretical exertions. The curious
circumstance that what was initially meant to model the field of a
``Massenpunkt'' displayed two singularities, one at the
``Schwarzschild radius'', and a second one for $r=0$, instead of
the single one appearing in Newtonian physics, suggested the idea
that one of them had to be spurious \footnote{Hilbert did not
share this opinion; however, he did not consider the singularities
displayed by his solution really worth of much study. He believed
in fact that the vacuum solutions with singularities were just
``an important mathematical tool for approximating characteristic
regular solutions''. To him, only the latter were capable to
represent reality in an immediate way. See \cite{Hilbert17}, page
70.}. Since the Kretschmann scalar happened to be finite at the
``Schwarzschild radius'', while it was infinite at $r=0$, the
conviction arose that the ``true'' singularity was the one at
$r=0$. Therefore the singularity displayed by the components of
the metric at $r=2m$ had to be a mere mathematical mishap, devoid
both of geometrical and of physical meaning. A reason had to be
given for the wrongdoing, and it was found in a presumed
inadequacy of the coordinate system at $r=2m$.\par The search thus
started for different coordinate systems that allowed to erase the
singular behaviour displayed by $g_{ik}$ at $r=2m$. Already in
1924 Eddington had unintentionally succeeded in the task
\cite{Eddington24} by rewriting the static ``Schwarzschild''
solution in stationary form through the introduction of what would
have been called the Eddington-Finkelstein coordinates
\footnote{Eddington's coordinate transformation was purposedly
retrieved \cite{Finkelstein58} by Finkelstein in 1958.}. In 1933
Lemaitre achieved the same result by rewriting \cite{Lemaitre33}
the ``Schwarzschild'' metric with cosmological term in
time-dependent form. Another solution to the problem was given in
1950 by Synge with a geometrically inspired paper \cite{Synge50}
that represents the now forgotten forerunner of the maximal
extensions \cite{Kruskal60}, \cite{Szekeres60} of the
``Schwarzschild'' metric obtained by Kruskal and Szekeres.\par All
these exertions entail coordinate transformations
${x'}^i=f^i(x^k)$ whose derivatives happen to be singular at the
``Schwarzschild'' radius in just the appropriate way for providing
a transformed metric that is regular there. One cannot help
noticing that the restriction to admissible coordinate
tranfrormations, which looked mandatory in the old papers, with
the lapse of the decades has become optional and dependent on
taste. In the time span that goes from Hilbert's paper
\cite{Hilbert17} to, say, the publication of Lichnerowicz' book
\cite{Lichnerowicz55} with his axioms inscribed in the first
chapter, transformations like the ones needed to efface the
``Schwarzschild'' singularity were simply disallowed.
\par
Nevertheless, the rule was violated here and there, and already in
Synge's paper one finds an explicit program of transgression,
since for the latter author ``it is precisely the non-regular
transformations which are interesting'' \cite{Synge50}. But the
value of scalars cannot be altered by any transformation, however
``interesting''. Therefore in all the alternative forms of the
``Schwarzschild'' metric mentioned in this section the singularity
in the metric components at the ``Schwarzschild'' radius is
canceled, but the norm $\alpha$ of the acceleration of the
hyperbolic motion on the invariantly specified Killing orbit
remains infinite at the position of the erased singularity. Since
$\alpha$ has a well defined physical meaning, an infinite value of
$\alpha$ in the middle of a manifold should not be so
light-heartedly overlooked: either this singularity should be
removed from that position, or a physical argument for its
existence there should be given.\par Already in 1916 Karl
Schwarzschild had deliberately sent the singularity in the
``Nullpunkt'', thus stipulating that there his idealised vacuum
model ceased to be physically meaningful, because the source of
the field had been attained.

\newpage
\bibliographystyle{amsplain}

\end{document}